\documentclass{LMCS}

\usepackage{amsmath,amssymb,xspace,epsfig,subfigure,latexsym,rotating,enumerate}

\newcommand{\comment}[1]{}

\newcommand{\printSideWays}[1]{{\begin{sideways}#1\end{sideways}}}

\newtheorem{theorem}{Theorem}
\newtheorem{proposition}{Proposition}

\newtheorem{definition}{Definition}
\newtheorem{remark}{Remark}

\usepackage{hyperref}
\comment{
\newcommand{\qed}{\mbox{} $\Box$}
\newenvironment{proof}{{\it Proof:}}{\qed}
}

\newcommand{\kw}[1]{{\bf #1}}

\newcommand{\true}{\kw{true}}
\newcommand{\false}{\kw{false}}

\newcommand{\sepreln}{\ensuremath{\bowtie}\xspace}

\newcommand{\maxA}{\ensuremath{a_{\max}}}
\newcommand{\maxB}{\ensuremath{b_{\max}}}
\newcommand{\amax}{\ensuremath{{a_{\max}}}\xspace}

\newcommand{\vect}[1]{\ensuremath{{\bf{#1}}}}
\newcommand{\myvec}[1]{\vect{#1}}

\newcommand{\fqfp}{\ensuremath{\Phi}\xspace}
\newcommand{\SolBound}{\ensuremath{d}\xspace}

\def\doi{1 (2:6) 2005}
\lmcsheading%
{\doi}
{1--26}
{}
{}
{Nov.~19, 2004}
{Dec.~19, 2005}
{}

\begin{document}

\title[Deciding Quantifier-Free Presburger Formulas]{Deciding Quantifier-Free Presburger Formulas Using Parameterized Solution Bounds}

\author[S.~A.~Seshia]{Sanjit A. Seshia\rsuper a}
\address{{\lsuper a}Department of Electrical Engineering and 
Computer Sciences,
University of California at Berkeley,\hfill\break
253 Cory Hall \#1770,
Berkeley, CA 94720, USA}
\email{sseshia@eecs.berkeley.edu}
\thanks{{\lsuper a}This work was done while the first author was at the School of Computer
Science, Carnegie Mellon University.}

\author[R.~E.~Bryant]{Randal E. Bryant\rsuper b}
\address{{\lsuper b}School of Computer Science,
Carnegie Mellon University,
5000 Forbes Avenue,\hfill\break
Pittsburgh, PA 15213, USA}
\email{Randy.Bryant@cs.cmu.edu}

%





\begin{abstract}
\noindent 
Given a formula in quantifier-free Presburger arithmetic, if it has a
satisfying solution, there is one whose size, measured in bits, is
polynomially bounded in the size of the formula. In this paper, we
consider a special class of quantifier-free Presburger formulas in
which most linear constraints are difference (separation) constraints,
and the non-difference constraints are sparse. This class has been
observed to commonly occur in software verification. We derive a new
solution bound in terms of parameters characterizing the sparseness of
linear constraints and the number of non-difference constraints, in
addition to traditional measures of formula size. In particular, we
show that the number of bits needed per integer variable is linear in
the number of non-difference constraints and logarithmic in the number
and size of non-zero coefficients in them, but is otherwise
independent of the total number of linear constraints in the
formula. The derived bound can be used in a decision procedure based
on instantiating integer variables over a finite domain and
translating the input quantifier-free Presburger formula to an
equi-satisfiable Boolean formula, which is then checked using a
Boolean satisfiability solver.  In addition to our main theoretical
result, we discuss several optimizations for deriving tighter bounds
in practice.  Empirical evidence indicates that our decision procedure
can greatly outperform other decision procedures.
\end{abstract}

\maketitle

\section{Introduction}
\label{sec:intro}

{\em Presburger arithmetic}~\cite{presburger-27} is the first-order
theory of the structure $\langle
\mathbb{N},0,1,\mbox{$\leqslant$},\mbox{$+$} \rangle$, where
$\mathbb{N}$ denotes the set of natural numbers.
The satisfiability problem for Presburger arithmetic is decidable, but of 
super-exponential worst-case complexity~\cite{fischer-siam74}.
Fortunately, for many applications, such as in program analysis (e.g.,~\cite{pugh-sc91})
and hardware verification (e.g.,~\cite{brinkmann-vlsi02}),
the quantifier-free fragment suffices.

A formula $\Phi$ in quantifier-free Presburger arithmetic (QFP) is constructed by combining linear 
constraints with Boolean operators ($\land$, $\lor$, $\neg$). Formally, the $i^{\text{th}}$ 
linear constraint is of the form
$\sum_{j=1}^n a_{i,j} x_j \geq b_i$,
where the coefficients and the constant terms are integer constants and the 
variables $x_1, x_2, \ldots, x_n$ are integer-valued\footnote{While Presburger
arithmetic is defined over $\mathbb{N}$, 
we interpret the variables over $\mathbb{Z}$ as it is general and more suitable for
applications. It is straightforward
to translate a formula with integer variables to one where variables are interpreted over $\mathbb{N}$,
and vice-versa, by adding (linearly many) additional variables or constraints.}. 
In this paper, we are concerned with the satisfiability problem for QFP, viz., that of finding 
a valuation of the variables such that $\Phi$ evaluates to \true{}.
The NP-hardness of this problem
follows from a straightforward encoding of the $3$SAT problem as a $0$-$1$ integer linear program.
That it is moreover in NP, and hence NP-complete, 
can be concluded from the result that integer linear programming is in 
NP~\cite{borosh-pams76,gathen-pams78,kannan-oor78,papadim-jacm81}.

Thus, if there is a satisfying solution to a QFP formula, 
there is one whose size, measured in bits, is polynomially bounded in the problem size. 
Problem size is traditionally 
measured in terms of the parameters $m$, $n$, $\log \maxA$, and $\log \maxB$, where
$m$ is the total number of constraints in the formula, $n$ is the number of variables, and
$\maxA = \max_{(i,j)} |a_{i,j}|$ and $\maxB = \max_i |b_i|$ are upper bounds on the absolute
values of coefficients and constant terms respectively.

The above result suggests the following approach to
checking the satisfiability of a QFP formula $\Phi$:
\begin{enumerate}
\item 
Compute the polynomial bound $S$ on solution size.
\item
Search for a satisfying solution to $\Phi$ in the bounded space $\{0, 1, \ldots, 2^S-1\}^n$.
\end{enumerate}
This approach has been successfully applied to highly restricted sub-classes of QFP, such
as {\em equality logic}~\cite{pnueli-cav99} and {\em difference logic}\footnote{{\em Difference logic} has
also been referred to as {\em separation logic} in the literature.}~\cite{bryant-cav02}, and is
termed as {\em finite instantiation} or the {\em small-domain encoding} approach.
The basic idea is to translate $\Phi$ to a Boolean formula by encoding
each integer variable as a vector of Boolean variables (a ``symbolic bit-vector'') 
of length $S$. 
The resulting Boolean formula is checked using a Boolean satisfiability (SAT) solver.
This approach leverages the dramatic advances in SAT solving made in recent years
(e.g.,~\cite{moskewicz-dac01,goldberg-date02}). 
It is straightforward to extend the approach to 
additionally handle the theory of uninterpreted functions and
equality, by using, for example, Ackermann's technique of eliminating function 
applications~\cite{ackermann-54}.

However, a na{\"\i}ve implementation of a decision procedure based on finite instantiation fails 
for QFP formulas encountered in practice. The problem is that 
the bound on solution size, $S$, is
$O(\log m + \log \maxB + m [\log m + \log \maxA])$. 
In particular, the presence of the $m \log m$ term means that,
for practical problems involving hundreds of linear constraints, the Boolean formulas
generated are likely to be too large to be decided by present-day SAT solvers.

\begin{table}[t]
\begin{center}
\begin{tabular} {|c|c|c|}
\hline
Project &   
Maximum Fraction of &
Maximum Width of a
\\
&
Non-Difference Constraints &
Non-Difference Constraint 
\\
\hline
Blast & 0.0255 & 6 \\ \hline
Magic & 0.0032 & 2 \\ \hline
MIT & 0.0087 & 3 \\ \hline
WiSA & 0.0091 & 4\\ \hline
\end{tabular}
\end{center}
\caption{{\bf{Linear Arithmetic Constraints in Software Verification are Mostly Difference Constraints.}}
For each software verification project, the maximum fraction of
non-difference constraints is shown, as well as the maximum width of a non-difference
constraint, where the 
maximum is taken over all formulas in the set.
The Blast formulas were generated from device drivers written in C, the Magic formulas
from an implementation of {\tt{openssl}} written in C,
the MIT formulas from Java programs, and 
the WiSA formulas were generated in the checking of format string vulnerabilities.
}
\label{tbl:mostly-sep}
\end{table}
In this paper, we explore the above finite instantiation-based approach to deciding
QFP formulas, but with a focus on formulas generated in software verification.
It has been observed, by us and others, that formulas from this domain have:
\begin{enumerate}[(1)]
\item {\em Mainly Difference Constraints:}
Of the $m$ constraints, $m-k$ are {\em difference} constraints, where $k \ll m$. 
Difference constraints, also called {\em separation} or {\em difference-bound} constraints, 
are of the form $ x_i - x_j \sepreln b_t $ or $ x_i \sepreln b_t $, 
where $b_t$ is an integer constant, and $\sepreln$ stands for a relational
symbol in the set $\{>, \geq, =, <, \leq\}$.
\item {\em Sparse Structure:}
The $k$ non-difference constraints are sparse, with at most $w$ variables per constraint,
where $w$ is ``small''. We will refer to $w$ as the {\em width} of the constraint.
\end{enumerate}
Pratt~\cite{pratt-77} observed that most inequalities generated in program verification
are difference constraints. More recently, 
the authors of the theorem prover Simplify observed in the
context of the Extended Static Checker for Java (ESC/Java) project that ``the inequalities that 
occur in program checking rarely involve more than two or three terms''~\cite{detlefs-tr03}.
We have performed a study of formulas generated in various recent software verification projects:
the Blast project at Berkeley~\cite{henzinger-popl02}, 
the Magic project at CMU~\cite{chaki-icse03}, 
the Wisconsin Safety Analyzer (WiSA) project~\cite{wisa-www},
and the software upgrade checking project at MIT~\cite{mccamant-fse03}. 
The results of this study, indicated in Table~\ref{tbl:mostly-sep}, support the afore-mentioned
observations regarding the ``sparse, mostly difference'' nature of constraints in QFP formulas.
To our knowledge, no previous decision procedure for QFP has attempted to
exploit this problem structure.

We make the following novel contributions in this paper:
\begin{enumerate}[$\bullet$]
\item
We derive bounds on solutions for QFP formulas, not only in terms of
the traditional parameters $m$, $n$, $\maxA$, and $\maxB$, but also in terms of
$k$ and $w$. In particular, we show that the worst-case number 
of bits required per integer variable is linear in
$k$, but only logarithmic in $w$. Unlike previously derived bounds, ours is 
not in terms of the total number of constraints $m$. 
\item
We use the derived bounds in a sound and complete
decision procedure for QFP based on finite instantiation, and present 
empirical evidence that
our method can greatly outperform other decision procedures.
\end{enumerate}

{\bf Related Work.} 
There has been much work on deciding quantifier-free Presburger arithmetic; we present
a brief discussion here and refer the
reader to a recent survey~\cite{ganesh-fmcad02} for more details. Recent techniques fall
into four categories: 
\begin{enumerate}[$\bullet$]
\item 
The first class comprises procedures targeted towards solving conjunctions of constraints,
with disjunctions handled by enumerating terms in a disjunctive normal form (DNF). 
Examples include the Omega test~\cite{pugh-sc91} (which is an extension of Fourier-Motzkin
elimination for integers) and solvers based on other integer linear
programming techniques.
The drawback of these methods is the need to enumerate the potentially exponentially many
terms in the DNF representation. Our work is targeted towards solving formulas with a
complicated Boolean structure, which often arise in verification applications.
\item 
The second set of methods attempt to remedy this problem by instead relying on modern
SAT solving strategies. The approach works as follows. A Boolean abstraction of the
QFP formula $\Phi$ is generated by replacing each linear constraint with a corresponding Boolean
variable. If the abstraction is unsatisfiable, then so is $\Phi$. If not, the satisfying
assignment (model) is checked for consistency with the theory of quantifier-free Presburger arithmetic,
using a ground decision procedure for conjunctions of linear constraints (i.e., a procedure
for checking feasibility of integer linear programs).
Assignments that are inconsistent are excluded from later consideration by adding a ``lemma''
to the Boolean abstraction. The process continues until either a consistent assignment is found, or
all (exponentially many) assignments have been explored.
Examples of decision procedures in this class that have some support for QFP 
include CVC~\cite{barrett-cav02,ganesh-tacas03} and ICS~\cite{demoura-cade02}.\footnote{The
general idea for combining a SAT solver with a linear programming engine originates 
in a paper by Wolfman and Weld~\cite{wolfman-ijcai99}.}
The ground decision procedures used by provers in this class employ a combination framework
such as the Nelson-Oppen architecture for cooperating decision procedures~\cite{nelson-toplas79}
or a Shostak-like combination method~\cite{shostak-jacm84,shankar-rta02}.
These methods are only defined for combining disjoint theories. 
In order to exploit the mostly-difference structure of a formula, one approach could be to
combine a decision procedure for a theory of difference constraints with one for a theory of
non-difference constraints, but this needs an extension of the combination methods 
that applies to these non-disjoint theories.
\item 
Strichman~\cite{strichman-fmcad02} presents SAT-based decision procedures for linear arithmetic (over the
rationals) and QFP. For QFP, the basic idea is to create a Boolean encoding of all the possible 
variable projection steps performed by the Omega test.
Since Fourier-Motzkin elimination (and therefore, the Omega test) 
has worst-case double-exponential complexity in both time and space~\cite{chandru-compjnl93}, 
this approach leads to a SAT problem that, in the worst-case, is doubly-exponential in the size
of the original formula and takes doubly-exponential time to generate. In contrast, in our approach
the SAT-encoding is polynomial in the size of the original formula, and is generated in polynomial
time.
\item
The final class of methods are based on automata theory 
(e.g.,~\cite{wolper-sas95,ganesh-fmcad02}). 
The basic idea in these methods is to construct a finite automaton corresponding to
the input QFP formula $\Phi$ such that the language accepted by the automaton
consists of the binary encodings of satisfying solutions of $\Phi$.
According to a recent experimental evaluation with other methods~\cite{ganesh-fmcad02},
these techniques are better than others at solving 
formulas with very large coefficients, but do not scale well with the number
of variables and constraints.\footnote{Note that automata-based techniques
can handle full Presburger arithmetic, not just the quantifier-free fragment.}
\end{enumerate}
The approach we present in this paper is distinct from the categories
mentioned above.
In particular, the following unique features differentiate it
from previous methods:
\begin{enumerate}[$\bullet$]
\item
It is the first finite instantiation method and the first tractable procedure for
translating a QFP formula to
SAT in a single step. The clear separation between the translation and the SAT solving
allows us to leverage future advances in SAT solving far more easily than other SAT-based
procedures.
\item
It is the first technique, to the best of our knowledge, that formally exploits the 
structure of formulas commonly encountered in software verification. 
\end{enumerate}
In addition to the above, the bounds we derive in this paper are also of independent
theoretical interest. For instance, they indicate that the solution bound does not
depend on the number of difference constraints.

{\bf Outline of the paper.}
The rest of this paper is organized as follows. In Section~\ref{sec:backgnd}, we discuss 
background material on bounds on satisfying solutions of integer linear programs. 
An integer
linear program (ILP) is a conjunction of linear constraints, and hence is a special kind of QFP formula.
The bounds for QFP follow directly from those for ILPs.
Our main theoretical results are presented in Section~\ref{sec:theory-main}.
Section~\ref{sec:sep} gives bounds for ILPs for the case of $k=0$,
when all constraints are difference constraints. In Section~\ref{sec:main}, we compute
a bound for ILPs for arbitrary $k$. In Section~\ref{sec:disjunctive}, we show how our 
results extend to arbitrary QFP formulas. 
Techniques for improving the bound in practice are discussed 
in Section~\ref{sec:enhancements}.
We report on experimental results in
Section~\ref{sec:results}, and conclude in Section~\ref{sec:concl}.

\section{Background}
\label{sec:backgnd}
In this section, we define the integer linear programming problem formally and state the
previous results on bounding satisfying solutions of ILPs. 
A more detailed discussion on the steps outlined in Section~\ref{sec:probdef} can be
found in reference books on ILP (e.g.~\cite{schrijver-86,papadim-steiglitz-82ch13}).

\subsection{Preliminaries}
\label{sec:probdef}
Consider a system of $m$ linear constraints in $n$ integer-valued variables:
\begin{equation}
A \vect{x} \geq b
\label{eqn:ineq-system-probdef}
\end{equation}
Here $A$ is an $m \times n$ matrix with integral entries, $b$ is a $m \times 1$ vector of integral entries,
and $\vect{x}$ is a $n \times 1$ vector of integer-valued variables.
A satisfying solution to system~(\ref{eqn:ineq-system-probdef}) is
an evaluation of $\vect{x}$ that satisfies~(\ref{eqn:ineq-system-probdef}).

In system~(\ref{eqn:ineq-system-probdef}), the entries in $\vect{x}$ can be negative.
We can constrain the variables to be non-negative by adding a dummy variable $x_0$ that refers to the 
``zero value,'' replacing each original variable $x_i$ by $x_i' - x_0$, and then adjusting the
coefficients in the matrix $A$ to get a new constraint matrix $A'$ and the following 
system:\footnote{Note that this procedure can increase the width of a constraint by $1$. 
The statistics in Table~\ref{tbl:mostly-sep} shows the width before this procedure is
applied, computed from constraints as they appear in the original formulas.}
\begin{equation}
\begin{split}
A' \vect{x}' & \geq b \\
\vect{x}' & \geq 0
\end{split}
\label{eqn:ineq-nz-system-probdef}
\end{equation}
Here the system has $n' = n+1$ variables, and $\vect{x}' = [x_1', x_2', \ldots, x_n', x_0]^T$.
$A'$ has the structure that $a_{i,j}' = a_{i,j}$ for $j = 1, 2, \ldots, n$ and 
$a_{i,n+1}' = - \sum_{j = 1}^n a_{i, j}$. Note that the last column of $A'$ is a linear
combination of the previous $n$ columns.
It is easy to show that
system~(\ref{eqn:ineq-system-probdef}) has a solution if and only if
system~(\ref{eqn:ineq-nz-system-probdef}) has one. 

Finally, adding surplus variables to the system, we can rewrite  
system~(\ref{eqn:ineq-nz-system-probdef}) as follows:
\begin{equation}
\begin{split}
A'' \vect{x}'' & = b \\
\vect{x}'' & \geq 0
\end{split}
\label{eqn:eq-system-probdef-primed}
\end{equation}
where $A'' = [A|-I_m]$ is an $m \times (n' + m)$ integer matrix formed by concatenating $A$ with the negation of
the $m \times m$ identity matrix $I_m$.\\
$\indent$ For convenience we will drop
the primes, referring to $A''$ and $\vect{x}''$ simply as $\vect{A}$ and $\vect{x}$. 
Rewriting system~(\ref{eqn:eq-system-probdef-primed}) thus, we get
\begin{equation}
\begin{split}
A \vect{x} & = b \\
\vect{x} & \geq 0
\end{split}
\label{eqn:eq-system-probdef}
\end{equation}
Hereafter we will mostly use the definition in~(\ref{eqn:eq-system-probdef}).

\begin{remark}
A solution to system~(\ref{eqn:eq-system-probdef}) also satisfies  
system~(\ref{eqn:ineq-nz-system-probdef}).
\label{remark:eq-ineq-nz-solutions}
\end{remark}

We next define two useful terms: {\em solution bound} and {\em enumeration bound}.

\begin{definition}
 Given a QFP formula $\fqfp$, a {\em solution bound}
 is an integer $\SolBound$ such that $\fqfp$ has an integer solution if and only if 
 it has an integer solution in 
 the $n$-dimensional hypercube $\prod_{i=1}^{n} [0, \SolBound]$.
\end{definition}

\begin{definition}
 Given a QFP formula $\fqfp$, an {\em enumeration bound}
 is an integer $\SolBound$ such that $\fqfp$ has an integer solution if and only if 
 it has an integer solution in
 the $n$-dimensional hypercube $\prod_{i=1}^{n} [-\SolBound, \SolBound]$.
 The interval $[-\SolBound, \SolBound]$ is termed as an {\em enumeration domain}.
\end{definition}

The following proposition is easily obtained. 
\begin{proposition}
A solution bound $\SolBound \geq 0$ for system~(\ref{eqn:ineq-nz-system-probdef}) is an 
enumeration bound for system~(\ref{eqn:ineq-system-probdef}).
\label{thm:sol-enum-bound}
\end{proposition}
\begin{proof}
Given a solution ${\myvec{x}'}^*$ to system~(\ref{eqn:ineq-nz-system-probdef}), we 
construct a solution $\myvec{x}^*$ to system~(\ref{eqn:ineq-system-probdef}) by setting 
$x_j^* = {x_j'}^* - x_0^*$. Since each ${x_j'}^*$ and $x_0^*$ are in $[0, \SolBound]$, 
$x_j^* \in [-\SolBound, \SolBound]$ for all $j$.  
\end{proof}

Similarly, if $d$ is an enumeration bound for 
system~(\ref{eqn:ineq-system-probdef}), then $2d$ is
a solution bound for system~(\ref{eqn:ineq-nz-system-probdef}).

Finally, we introduce symbols $\maxA$ and $\maxB$ with the following associated
meanings: $\maxA = \max_{(i,j)} |a_{i,j}|$ and $\maxB = \max_t |b_t|$. 
In words, $\maxA$ and $\maxB$ are tight upper bounds on the absolute
values of entries of $A$ and $b$ respectively.

\subsection{Previous Results}
\label{sec:prev-results}

The results of this paper build on results obtained by
Borosh, Treybig, and Flahive~\cite{borosh-pams76,borosh-dismath86} 
on bounding the solutions of systems of the form~(\ref{eqn:eq-system-probdef}).
We state their result in the following theorem:
\begin{theorem} 
Consider the augmented matrix $[A|b]$ of dimension $m \times (n'+m+1)$. Let $\Delta$ be the maximum
of the absolute values of all minors of this augmented matrix. Then, the system~(\ref{eqn:eq-system-probdef})
has a satisfying solution if and only if it has one with all entries bounded by $(n+2) \Delta$.\qed
\label{thm:borosh-treybig}
\end{theorem}
However, note that the determinant of a matrix can be more than exponential in the dimension of the 
matrix~\cite{brenner-amm72}.
In the case of the Borosh-Flahive-Treybig
result, it means that $\Delta$ can be as large as $\frac{{\mu}^m (m+1)^{(m+1)/2}}{2^m}$, where
$\mu = \max(\maxA,\maxB)$.

Papadimitriou~\cite{papadim-jacm81,papadim-steiglitz-82ch13} also gives a 
bound of similar size, stated in the following theorem:
\begin{theorem}
If the ILP of~(\ref{eqn:eq-system-probdef}) has a satisfying solution, then it has a satisfying
solution where all entries in the solution vector are bounded by $(n'+m) (1 + \maxB) (m \, \maxA)^{2m+3}$.\qed
\label{thm:papadim-steiglitz}
\end{theorem}
Papadimitriou's bound implies that we need 
$O(\log m + \log \maxB + m [\log m + \log \maxA])$ bits to encode each variable (assuming $n' = O(m)$).
The Borosh-Flahive-Treybig bound implies needing $O(m [\log m + \log \mu])$ bits per variable, which is
of the same order.

\section{Main Theoretical Results}
\label{sec:theory-main}

\subsection{Bounds for a System of Difference Constraints}
\label{sec:sep}

Let us first consider computing solution bounds for an ILP for the case where $k = 0$, 
i.e., system~(\ref{eqn:eq-system-probdef}) comprises only of difference constraints. 

In this case, the left-hand side of each equation comprises exactly three variables: two
variables $x_i$ and $x_j$ where $0 \leq i,j \leq n$ and one surplus variable $x_l$ where
$n+1 \leq l \leq n+m$.
The $t^{\text{th}}$ equation in the system is of the form $x_i - x_j - x_l = b_t$.

As we noted in Section~\ref{sec:probdef}, the matrix $A$ can be written as $[A_o | -I_m]$
where $A_o$ comprises the first $n'=n+1$ columns, and $I_m$ is the $m \times m$ identity matrix.

The important property of $A_o$ is that each row has exactly one $+1$ entry and exactly one $-1$ entry,
with all other entries $0$. Thus, $A_o^T$ can be interpreted as the node-arc incidence matrix of
a directed graph. 
Therefore, $A_o^T$ is {\em totally unimodular} (TUM), i.e., every square submatrix of $A_o^T$ has
determinant in $\{0, -1, +1\}$~\cite{papadim-steiglitz-82ch13}.
Therefore, $A_o$ is TUM, and so is $A = [A_o | -I_m]$.

Now, let us consider using the Borosh-Flahive-Treybig bound stated in Theorem~\ref{thm:borosh-treybig}. 
This bound is stated in terms of the minors of the matrix $[A | b]$. 
For the special case
of this section, we have the following bound on the size of any minor:
\begin{theorem}
The absolute value of any minor of $[ A | b ]$ is bounded above by $s \, \maxB$, where
$s = \min(n+1, m)$.
\label{thm:sep-bound}
\end{theorem}

\proof
Consider any minor $M$ of $[A | b]$. Let $r$ be the order of $M$.

If the minor is obtained by deleting the last column (corresponding to $b$), then it is a
minor of $A$, and its value is in $\{0,-1,+1\}$ since $A$ is TUM. Thus, the bound of $s \, \maxB$
is attained for any non-trivial minor with $s \geq 1$ and $\maxB \geq 1$.

Suppose the $b$ column is not deleted. 

First, note that the matrix $A$ is of the form $[A_o | -I_m]$ where
the rank of $A_o$ is at most $s' = \min(n, m)$. This is because $A_o$ has dimensions $m \times n+1$,
and the last column of $A_o$, corresponding to the variable $x_0$, is a linear combination of the previous 
$n$ columns. (Refer to the construction of system~(\ref{eqn:ineq-nz-system-probdef}) from 
system~(\ref{eqn:ineq-system-probdef}).)

Next, suppose the sub-matrix corresponding to $M$ comprises $p$ columns from the $-I_m$ part,
$r-p-1$ columns from the $A_o$ part, and the one column corresponding to $b$. Since permuting the
rows and columns of $M$ does not change its absolute value, we can permute the
rows of $M$ and the columns corresponding to the $-I_m$ part to get the corresponding
sub-matrix in the following form:
$$
\begin{bmatrix}
&  & \vline & 0 & \ldots & 0 & -1 & \vline & b_{t_1} \\ 
&  & \vline & 0 & \ldots & -1 &  0 & \vline & b_{t_2} \\ 
& A_o & \vline & \vdots & \cdots & \vdots &  \vdots & \vline & \vdots \\ 
& \mbox{part} & \vline & -1 & \ldots & 0 &  0 & \vline & b_{t_p} \\ 
& & \vline &  0 & \ldots & 0 &  0 & \vline & b_{t_{p+1}} \\ 
& & \vline & \vdots & \cdots & \vdots &  \vdots & \vline & \vdots \\ 
& & \vline &  0 & \ldots & 0 &  0 & \vline & b_{t_r} \\ 
\end{bmatrix}
$$

Expanding $M$ along the last column, we get
$$ |M|  =  | b_{t_1} M_1 - b_{t_2} M_2 + b_{t_3} M_3 - \ldots (-1)^{r-1} b_{t_r} M_r | $$
where each $M_i$ is a minor corresponding to a submatrix of $A$. 

However, notice that $M_i = 0$ for all $1 \leq i \leq p$, since each of those minors have an entire
column (from the $-I_m$ part) equal to $0$. Therefore, we can reduce the right-hand side to 
the sum of $r-p$ terms:
$$ |M|  \leq  | b_{t_{p+1}} M_{p+1} |  + | b_{t_{p+2}} M_{p+2} |  + \ldots | b_{t_r} M_r | $$

Notice that, so far, we have not made use of the special structure of $A$. 

Now, observing that $A$ is TUM, $|M_i| \leq 1$ for all $i$. 
$$ |M|  \leq  | b_{t_{p+1}} | + | b_{t_{p+2}} | + \ldots +  | b_{t_r} | $$

For all $i$, $|b_{t_i}| \leq \maxB$. Further, since each non-zero $M_i$ can be
of order at most $s'$, $r-p \leq s = min(s'+1, m)$.\footnote{We
use $s'+1$ and not $s'$ to account for the case where $p=0$. The minimum with $m$ is taken
because $s'+1$ can exceed $m$ but $b$ has only $m$ elements.} Therefore, we get
$$
\hbox to182 pt{\hfil}
|M| \leq s \, \maxB 
\hbox to182 pt{\hfil}\qEd
$$

Using the terminology of Theorem~\ref{thm:borosh-treybig}, we have $\Delta \leq s \, \maxB$.
Thus, the bound in this case is $(n+2) \, s \, \maxB$. 

Thus, $S$, the bound on the number of bits per variable, is
$$\lceil \log (n+2) + \log s + \log \maxB \rceil$$
Formulas generated from verification problems tend to be overconstrained,
so we assume $n < m$. Thus, $s = n+1$, and the bound reduces to
$O(\log n + \log \maxB)$ bits per variable. 

{\em Remark:}
The only property of the $A$ matrix that the proof of 
Theorem~\ref{thm:sep-bound} relies on is the totally unimodular (TUM) property. 
Thus, Theorem~\ref{thm:sep-bound}
would also apply to any system of linear constraints whose coefficient matrix is
TUM. Examples of such matrices include {\em interval} matrices, or more generally
{\em network} matrices. Note that the TUM property can be tested for in polynomial 
time~\cite{schrijver-86}.

\subsection{Bounds for a Sparse System of Mainly Difference Constraints}
\label{sec:main}

We now consider the general case for ILPs, where we have $k$ non-difference constraints, each 
referring to at most $w$ variables.

Without loss of generality, we can reorder the rows of matrix $A$ so that the $k$
non-difference constraints are the top $k$ rows, and the difference constraints are the bottom
$m-k$ rows. Reordering the rows of $A$ can only change the sign of any minor of $[A | b]$, not
the absolute value.
Thus, the matrix $[A | b]$ can be put into the following form:
$$
\begin{bmatrix}
& A_1 & \vline & & \vline & b_1 \\ \cline{1-2} 
& & \vline & -I_m & \vline & b_2 \\
& A_2 & \vline & & \vline & \vdots \\
& & \vline &  & \vline & b_m \\
\end{bmatrix}
$$
Here, $A_1$ is a $k \times n+1$ dimensional matrix corresponding to the non-difference constraints,
$A_2$ is a $m-k \times n+1$ dimensional matrix with the difference constraints, $I_m$ is the
$m \times m$ identity corresponding to the surplus variables, and the last column is the vector $b$.

For ease of presentation, we will assume in 
the rest of Sections~\ref{sec:main} and~\ref{sec:disjunctive} that $k \leq n+1$. 
We will revisit this assumption at the end of Section~\ref{sec:theory-main}.

The matrix composed of $A_1$ and $A_2$ will be referred to, as before, as $A_o$.
Note that each row of $A_1$ has at most $w$ non-zero entries, and each row of $A_2$ has exactly
one $+1$ and one $-1$ with the remaining entries $0$. Thus, $A_2$ is TUM.

We prove the following theorem:
\begin{theorem}
The absolute value of any minor of $[ A | b ]$ is bounded above by $s \, \maxB \, (\maxA \, w)^k$,
where $s = \min(n+1, m)$. 
\label{thm:main-bound}
\end{theorem}
\begin{proof}

Consider any minor $M$ of $[ A | b ]$, and let $r$ be its order. 

As in Theorem~\ref{thm:sep-bound}, if $M$ includes $p$ columns from the $- I_m$ part of $A$, then 
we can infer that $r - p \leq s$. 
(Our proof of this property in Theorem~\ref{thm:sep-bound} 
made no assumptions on the form of $A_o$.)

If $M$ includes the last column $b$, then as in the proof of Theorem~\ref{thm:sep-bound}, we can conclude
that 
\begin{equation}
|M| \, \leq \, (r - p) \, \maxB \; [ \max_{j=1}^{r} | M_j | ] 
\label{eqn:main-bound-minor}
\end{equation}
where $M_j$ is a minor of $A_o$.

If $M$ does not include $b$, then it is a minor of $A$.
Without loss of generality, we can assume that $M$ does not include a column from 
the $-I_m$ part of $A$, since such columns only contribute to the sign of the determinant.

So, let us consider bounding a minor $M_j$ of $A_o$ of order $r$ 
(or $r-1$, if $M$ includes the $b$ column). 

Since $A_o = \begin{bmatrix} \frac{A_1}{A_2} \end{bmatrix}$, consider expanding $M_j$, using the
standard determinant expansion by minors along the top
$k$ rows corresponding to non-difference constraints. 
Each term in the expansion is (up to a sign) the product of at most $k$ entries from
the $A_1$ portion, one from each row, 
and a minor from $A_2$. Since $A_2$ is TUM, each product term is bounded in
absolute value by $\maxA^k$. Furthermore, there can be at most $w^k$ non-zero terms in the expansion, 
since each non-zero product term is obtained by choosing one non-zero element from each of the rows of the $A_1$
portion of $M_j$, and this can be done in at most $w^k$ ways.

Therefore, $|M_j|$ is bounded by $(\maxA \, w)^k$. Combining this with the inequality~(\ref{eqn:main-bound-minor}),
and since $r - p \leq s$, we get
$$ | M | \, \leq \, s \, \maxB \; (\maxA \, w)^k $$
which is what we set out to prove.
\end{proof}

Thus, we conclude that 
$\Delta \leq s \, \maxB (\maxA \, w)^k$, where $s = \min(n+1,m)$. From 
Theorems~\ref{thm:borosh-treybig} and~\ref{thm:main-bound}, and 
Remark~\ref{remark:eq-ineq-nz-solutions},
we obtain the following theorem:
\begin{theorem}
A solution bound for the system~(\ref{eqn:ineq-nz-system-probdef}) is 
$$
(n+2) \Delta = (n+2) \cdot s \cdot \maxB \cdot (\maxA \, w)^k
$$
\label{thm:sol-bound-conjunction}
\end{theorem}
Thus, the solution size $S$ is
$$\lceil \log (n+2) + \log s + \log \maxB + k (\log \maxA + \log w) \rceil$$

We make the following observations about the bound derived above, assuming
as before, that $n < m$, and so $s = n+1$:
\begin{enumerate}[$\bullet$]
\item {\em Dependence on Parameters:} 
We observe that the bound is linear in $k$,
logarithmic in $\maxA$, $w$, $n$, and $\maxB$. In particular, the bound is
not in terms of the total number of linear constraints, $m$.

\item {\em Worst-case Asymptotic Growth}:
In the worst case, $k = m$, $w = n+1$, and $n = O(m)$, and we get the 
$O(\log m + \log \maxB + m [\log m + \log \maxA])$ bound of Papadimitriou.

\item {\em Typical-case Asymptotic Growth:}
As observed in our study of formulas from software verification, 
$w$ is typically a small constant, so the number of bits needed per variable
is $O(\log n + \log \maxB + k \log \maxA + k)$. 
In many cases, $\maxA$ and $k$ are also bounded by a small constant. Thus, $S$ 
is typically $O(\log n + \log \maxB)$. This reduces the search
space by an exponential factor over using the bound expressed in terms of $m$.

\item {\em Representing Non-difference Constraints:} There are many ways to represent
non-difference constraints and these have an impact on the bound we derive.
In particular, it is possible to transform a system of non-difference
constraints to one with at most three variables per constraint.
For example, the linear constraint $x_1 + x_2 + x_3 + x_4 = x_5$ can
be rewritten as:
\begin{eqnarray*}
x_1 + x_1' & = & x_5 \\
x_2 + x_2' & = & x_1' \\
x_3 + x_4 & = & x_2' 
\end{eqnarray*}
For the original representation, $k = 1$ and $w = 5$, while for the
new representation $k = 3$ and $w = 3$. Since our bound is linear
in $k$ and logarithmic in $w$, the original representation would
yield a tighter bound.

Similarly, one can eliminate variables with coefficients greater than $1$ in
absolute value by introducing new variables; e.g., $2 x$ is represented
as $x + x'$ with an additional difference constraint $x = x'$. 
This can be used to adjust $w$, $\maxA$, and $n$ so that the overall bound
is reduced.
\end{enumerate}

The derived bound only yields benefits in the case when 
the system has few non-difference constraints which themselves are sparse. In this case,
we can instantiate variables over a finite domain that is much smaller than that obtained
without making any assumptions on the structure of the system.

Finally, from Proposition~\ref{thm:sol-enum-bound} and
Theorem~\ref{thm:sol-bound-conjunction}, 
we obtain an enumeration bound for system~(\ref{eqn:ineq-system-probdef}):
\begin{theorem}
An enumeration bound for system~(\ref{eqn:ineq-system-probdef}) is 
$$ (n+2) \cdot s \cdot \maxB \cdot (\maxA \, w)^k $$
\label{thm:enum-bound-conjunction}
\end{theorem}
Note that the values of $\maxA$ and $w$ in the statement of Theorem~\ref{thm:enum-bound-conjunction}
are those for system~(\ref{eqn:ineq-nz-system-probdef}).

\subsection{Bounds for Arbitrary Quantifier-Free Presburger Formulas}
\label{sec:disjunctive}

We now return to the original goal of this paper, that of finding a
solution bound for an arbitrary QFP formula
$\Phi$. 

Suppose that $\Phi$ has $m$ linear constraints $\phi_1, \phi_2, \ldots, \phi_m$, 
of which $m - k$ are difference constraints, and
$n$ variables $x_1, x_2, \ldots, x_n$.
As before, we assume that each non-difference
constraint has at most $w$ variables, $\maxA$ is the maximum over the absolute values of
coefficients $a_{i,j}$ of variables, and $\maxB$ is the maximum over the absolute values of constants
$b_i$ appearing in the constraints. Furthermore, let us assume that the
zero variable (used in transforming system~\ref{eqn:ineq-system-probdef} to 
system~\ref{eqn:ineq-nz-system-probdef}) have already been introduced into the constraints.

We prove the following theorem.
\begin{theorem}
If $\Phi$ is satisfiable, there is a solution to $\Phi$ that is bounded by $(n+2) \Delta$ where
$$\Delta = s \, (\maxB + 1) \, (\maxA \, w)^k$$
and $s = \min(n+1, m)$. 
\label{thm:disj-bound}
\end{theorem}
\begin{proof}
Let $\sigma$ be a (concrete) model of $\Phi$.
Let $m'$ constraints,
$\phi_{i_1}, \phi_{i_2}, \ldots, \phi_{i_{m'}}$, 
evaluate to $\true$ under $\sigma$, the rest evaluating to $\false$. Let $A' = [a_{i,j}]$ be a
$m' \times n$ matrix in which each row comprises the coefficients of variables 
$x_1, x_2, \ldots, x_n$ in a constraint $\phi_{i_k}$, $1 \leq k \leq m'$.
Thus,  $A' = [a_{i,j}]$ where $i \in \{i_1, \ldots, i_{m'}\}$.

Now consider a constraint $\phi_{i_k}$ where $k > m'$, that evaluates to $\false$ under $\sigma$.
$\phi_{i_k}$ is the inequality
$$ \sum_{j=1}^{n} a_{i_k,j} x_j \geq b_{i_k} $$
Then $\sigma$ satisfies $\neg \phi_{i_k}$ which is the inequality
$$ \sum_{j=1}^{n} a_{i_k,j} x_j < b_{i_k} $$
or equivalently, 
$$ \sum_{j=1}^{n} - a_{i_k,j} x_j \geq - b_{i_k} + 1 $$
Let $A''$ be a $(m-m') \times n$ matrix corresponding to the coefficients of variables in constraints 
$\neg \phi_{i_{m'+1}}$, $\neg \phi_{i_{m'+2}}$, $\ldots$, $\neg \phi_{i_{m}}$.
Thus, $A'' = [- a_{i,j}]$ where $i \in \{i_{m'+1}, \ldots, i_{m}\}$.

Finally, let 
$b = [b_{i_1}, b_{i_2}, \ldots, b_{i_{m'}}, - b_{i_{m'+1}} + 1, - b_{i_{m'+2}} + 1, \ldots, - b_{i_m} + 1]^T$

Clearly, $\sigma$ is a satisfying solution to the ILP given by
\begin{equation}
\begin{bmatrix} & A' & \\ \cline{2-2} & A'' & \end{bmatrix} \vect{x} \geq b  
\label{eqn:ilp-disj-thm}
\end{equation}
Also, if the system~(\ref{eqn:ilp-disj-thm}) has a satisfying solution then $\Phi$ is satisfied by that solution.
Thus, $\Phi$ and the system~(\ref{eqn:ilp-disj-thm}) are equi-satisfiable, for every possible
system~(\ref{eqn:ilp-disj-thm}) we construct in the manner described above.

By Theorems~\ref{thm:borosh-treybig} and~\ref{thm:main-bound}, we can conclude that if 
system~(\ref{eqn:ilp-disj-thm}) has a satisfying solution, it has one bounded by $(n+2) \Delta$ where
$$ \Delta = s \, (\maxB + 1) \, (\maxA \, w)^k$$
and $s = \min(n+1, m)$. 
Moreover, this bound works for every possible system~(\ref{eqn:ilp-disj-thm}).

Therefore, if $\Phi$ has a satisfying solution, it has one bounded by $(n+2) \Delta$.
\end{proof}

Thus, to generate the Boolean encoding of the starting QFP formula, we must encode each integer
variable as a symbolic bit-vector of length $S$ given by 
$$
S = \lceil \log[(n+2) \Delta] \rceil 
  = \lceil \log (n+2) + \log s + \log (\maxB + 1) + k (\log \maxA + \log w) \rceil
$$

\begin{remark}
If the zero variable is not introduced into the formula $\fqfp$, we can
search for solutions in $\prod_{i=1}^{n} [-\SolBound, \SolBound]$, where 
$\SolBound = (n+2) \Delta$. As noted earlier,
values of $\maxA$ and $w$ used in computing $\Delta$ 
are those obtained after introducing the zero variable.
\end{remark}

\begin{remark}
In Section~\ref{sec:main}, we assumed, for ease of presentation, 
that $k \leq n+1$. If this does not hold, we can simply replace
$k$ in the results of Sections~\ref{sec:main} and~\ref{sec:disjunctive} by
$\min(k, n+1)$. This is because the dimension of the minor $M_j$ of $A_o$ 
(mentioned in the proof of Theorem~\ref{thm:main-bound}) is limited by $n+1$.
\end{remark}

We conclude this section by summarizing the symbols used to represent formula
parameters and the quantities derived therefrom. For easy reference, they
are listed in Table~\ref{tbl:symbols}.
\begin{table}[h]
\begin{center}
\begin{tabular}{|c|l|}
\hline
{\it Symbol} & {\it Meaning} \\ \hline
\hline
$n$ & Number of variables \\ \hline
$m$ & Number of constraints \\ \hline
$\maxB$ & Maximum constant term \\ \hline
$\maxA$ & Maximum variable coefficient \\ \hline
$k$ & Number of non-difference constraints \\ \hline
$w$ & Maximum number of non-zero coefficients in any constraint \\  \hline
$s$ & $\min(n+1, m)$ \\  \hline
$\Delta$ & $s \cdot (\maxB + 1) \cdot (\maxA \, w)^k$ \\ \hline
$S$ & $\lceil \, \log[(n+2) \Delta] \, \rceil$ \\ \hline
\end{tabular}
\end{center}
\caption{{\bf{Parameters and Derived Quantities.}}}
\label{tbl:symbols}
\end{table}

\section{Improvements}
\label{sec:enhancements}

The bounds we derived in the preceding section are conservative. For a particular
problem instance, the size of minors can be far smaller than the bound we 
computed. However, this cannot be directly exploited by enumerating minors,
since the number of minors
grows exponentially with the dimensions of the constraint matrix.
Also, there is a special case under which
one can improve the $(n+2) \Delta$ bound. 
If all the constraints are originally equalities and the system of constraints has full rank, 
a bound of $\Delta$ suffices~\cite{borosh-pams89}. However, in our experience, even if 
the linear constraints are all equalities, they still tend to be linearly dependent. Thus,
we have not been able to make use of this special case result.

Fortunately, there are other techniques for improving the solution bound that we have
found to be fairly useful in practice.
These include theoretical improvements as well as heuristics that are useful in practice.
We describe these methods in this section.

\subsection{Variable Classes}
\label{sec:var-classes}

So far, we have used a single bit-vector length for
all integer variables appearing in the formula $\Phi$. This is overly conservative.
In general, we can partition the set of variables into classes such that two variables
are placed in the same class if there is a constraint in which they both appear with non-zero 
coefficients. 
Note, moreover, that this partitioning optimization can be 
performed {\em before} adding the ``zero'' variable
$x_0$. A different zero variable is then used for each variable class.
For each class, we separately compute parameters $n$, $k$, $\maxB$, $\maxA$, and $w$,
resulting in a separately computed bit-vector length for each class.

For example, consider the formula 
$$
x_1 + x_2 \geq 1 
\; \land \; 
\bigl( x_2 - x_3 \geq 0 \; \lor \; x_4 - x_5 \geq 0 \bigr)
$$
In this case, variables $x_1$, $x_2$, and $x_3$ fall into one class, while $x_4$ and $x_5$
will be put into a different class.

The correctness of this partitioning optimization follows from a reduction to
ILP as performed in the proof of Theorem~\ref{thm:disj-bound}, along with 
the following two observations: 
\begin{enumerate}[$\bullet$]
\item By construction, different variable classes share neither variables nor constraints.
\item A different zero variable can be introduced for each class because that transformation 
preserves solutions in the same way as the transformation from system~(\ref{eqn:ineq-system-probdef})
to system~(\ref{eqn:ineq-nz-system-probdef}) does.
\item A satisfying solution to a system of ILPs, no two of 
which share a variable, can be obtained by 
solving them independently and concatenating the solutions.
\end{enumerate}

\subsection{Tighter Bounds for Special Constraint Classes}
\label{sec:spl-case-bounds}

Consider specializing the solution bound of Section~\ref{sec:disjunctive}
to the special cases of equality logic and difference logic. 
(An equality logic formula only has constraints of the form $x_i = x_j$.)

For equality logic, $k = 0$, and $\maxB = 0$.
Thus, our bound specializes to $(n+2) \cdot s$, which, assuming $n < m$, is $O(n^2)$. 
For separation logic too, $k = 0$. 
This yields a bound of $(n+2) \cdot s \cdot (\maxB + 1)$.

However, both of these bounds are too conservative. 

For an equality logic formula with $n$ variables,
it is well-known that a solution bound of $n$ suffices
to decide the satisfiability of the formula. 

Similarly, if the formula is in difference logic, a solution bound of
$\min(n,m) \cdot (\maxB + 1)$ suffices. We sketch the proof of this result here,
omitting details. The proof is based on a graph-theoretic view of difference-bound
constraints, with each variable corresponding to a vertex, and a constraint
$x_i \geq x_j + b_t$ corresponding to an edge from $x_i$ to $x_j$ of weight $b_t$.
(The graph is constructed after first putting the formula into negation normal form;
see the paper by Strichman et al.~\cite{strichman-cav02} for details on graph
construction.)
A satisfying assignment is an assignment of integers to vertices such that the
graph has no positive cycles. Now note that, in this graph, the longest path is
of length $\min(n,m) \cdot (\maxB + 1)$, since there are $n+1$ vertices in the graph
(including that for the zero variable) and the weight of any edge is at most
$\maxB + 1$. Thus, if there is a satisfying assignment, there is one in
which the separation between the minimum and maximum integer value does not
exceed $\min(n,m) \cdot (\maxB + 1)$. This concludes the proof sketch.

Clearly, if the formula is purely in equality logic or purely in difference
logic, we can use the tighter bounds for the appropriate logic. However, the 
optimization of computing variable classes (presented in Section~\ref{sec:var-classes}) 
allows us to exploit the tighter bounds even if
the overall formula is not in equality logic or difference logic:
The tighter bounds can be used for encoding variables
in variable classes that comprise purely equality or purely difference constraints.
The correctness of this optimization follows for the same reasons as that
of the original variable class partitioning optimization.

\subsection{Dealing with Large Coefficients and Widths}
\label{sec:large-coeff}

In the expression for $S$, the term involving $\maxA$ (and $w$) is
multiplied by a factor of $k$.
Thus, any increase in $\log \maxA$ gets amplified by a factor of $k$. It is therefore
useful, in practice, to more carefully model the dependence of $S$ on coefficients.
We present two techniques to alleviate the problem of dealing with large coefficients.
These techniques also apply to dealing with large constraint widths.

\subsubsection{An $n^k$-fold reduction}
\label{sec:coeff-zero-var-opt}

The coefficient of the zero variable $x_0$ has, so far, been used in computing
$\maxA$. We will now show that we can ignore this coefficient, and also ignore
any contribution of $x_0$ to the width $w$.
This optimization can result in a reduction of up to a factor of $n^k$ in the
solution bound $\SolBound$.

The largest reduction occurs when, in the original formula, 
we have a constraint of the form $\sum_j a_i x_j \geq b_i$, where $a_i$ is
the largest coefficient in absolute value. After adding the zero variable,
this constraint is transformed to $(\sum_j a_i x_j) - (n \cdot a_i) x_0 \geq b_i$.
Thus, $\maxA$ now equals $n \cdot a_i$, a factor of $n$ times greater than in the original formula.

Let us revisit the transformation performed in Section~\ref{sec:probdef} to convert
system~(\ref{eqn:ineq-system-probdef}) to system~(\ref{eqn:ineq-nz-system-probdef}).
A different and commonly-used transformation to non-negative variables is to 
write each $x_j$ as $x_j^+ - x_j^-$, where $x_j^+, x_j^- \geq 0$ for all $j$.
Let the resulting system be referred to as system~(\ref{eqn:ineq-nz-system-probdef}').
Let us assume that this different transformation is used in place of the original one 
that generates system~(\ref{eqn:ineq-nz-system-probdef}),
leaving all successive transformations the same.

Now, consider the form of the matrix $[A | b]$, as used in Section~\ref{sec:main},
reproduced below:
$$
\begin{bmatrix}
& A_1 & \vline & & \vline & b_1 \\ \cline{1-2} 
& & \vline & -I_m & \vline & b_2 \\
& A_2 & \vline & & \vline & \vdots \\
& & \vline &  & \vline & b_m \\
\end{bmatrix}
$$
With the new transformation method, $A_1$ is a $k \times 2n$ dimensional 
matrix corresponding to the non-difference constraints,
$A_2$ is a $(m-k) \times 2n$ dimensional matrix with the difference constraints, 
$I_m$ is the $m \times m$ identity corresponding to the surplus variables, 
and the last column is the vector $b$.

Importantly, note that $A_2$ is {\em still} totally unimodular and the ranks of $A_1$
and $A_2$ are the same as they were with the use of the single zero variable $x_0$.
This is because any non-singular sub-matrix 
of $A_o$ must include exactly one of the columns corresponding to $x_i^+$ and $x_i^-$,
since they are negations of each other.
Therefore, the values of $w$ and $\maxA$ used in the proof of Theorem~\ref{thm:main-bound}
are those for the
system~(\ref{eqn:ineq-system-probdef}).

Thus, if we use the transformation method of replacing $x_i$ with $x_j^+ - x_j^-$,
the values of $w$ and $\maxA$ used in the statement of 
Theorem~\ref{thm:main-bound} are those for the system~(\ref{eqn:ineq-system-probdef}).

Note, however, that by replacing $x_i$ with $x_j^+ - x_j^-$, the number of variables
in the problem doubles, and in particular, the number of input variables in the
SAT-encoding is doubled. This is rather undesirable.

Fortunately, there are two solutions that avoid the doubling of variables at the minor
cost of only $1$ extra bit per variable.
\begin{enumerate}[(1)]
\item
The first solution is based on  
the following proposition that mirrors Proposition~\ref{thm:sol-enum-bound}.
\begin{proposition}
A solution bound $\SolBound \geq 0$ for system~(\ref{eqn:ineq-nz-system-probdef}') is an 
enumeration bound for system~(\ref{eqn:ineq-system-probdef}).
\label{thm:sol-enum-bound2}
\end{proposition}
\begin{proof}
Given a solution ${\myvec{x}'}^*$ within the solution bound $\SolBound$ to system~(\ref{eqn:ineq-nz-system-probdef}'), 
we construct a solution $\myvec{x}^*$ to system~(\ref{eqn:ineq-system-probdef}) by setting 
$x_j^* = {x_j^+}^* - {x_j^-}^*$. Clearly, 
$x_j^* \in [-\SolBound, \SolBound]$ for all $j$.  
\end{proof}

Thus, we can restrict our search to the hypercube $\prod_{i=1}^{n} [-\SolBound, \SolBound]$, where
the solution bound $\SolBound$ is computed using the values of $w$ and $\maxA$ for the
system~(\ref{eqn:ineq-system-probdef}).

\item
The second solution uses the following proposition showing that we can use the technique of
adding a zero variable $x_0$ and the values of $w$ and $\maxA$ for the
system~(\ref{eqn:ineq-system-probdef}), while paying only a minor penalty of
$1$ extra bit per variable.
\begin{proposition}
Suppose $\SolBound \geq 0$ is a solution bound
such that system~(\ref{eqn:ineq-nz-system-probdef}') has a solution in $[0, \SolBound]$
iff system~(\ref{eqn:ineq-system-probdef}) is feasible.
Then, system~(\ref{eqn:ineq-nz-system-probdef}) has a solution in $[0, 2 \SolBound]$ iff
system~(\ref{eqn:ineq-nz-system-probdef}') has a solution in $[0, \SolBound]$.
\end{proposition}
\begin{proof}

{\it (if part)}:
Suppose system~(\ref{eqn:ineq-nz-system-probdef}') has a solution in $[0, \SolBound]$; i.e.,
$x_j^+, x_j^- \in [0, \SolBound]$ for all $j$.
Then, we construct a satisfying assignment to system~(\ref{eqn:ineq-nz-system-probdef})
as follows:
\begin{itemize}
\item $x_0$ is assigned the value $\max_j x_j^-$.
\item $x_j$, for $j > 0$, is assigned the value $x_j^+ + (x_0 - x_j^-)$.
\end{itemize}
Since $0 \leq (x_0 - x_j^-) \leq \SolBound$, we can conclude that $0 \leq x_j \leq 2 \SolBound$ for all $j$.
It is easy to see that the resulting assignment satisfies 
system~(\ref{eqn:ineq-nz-system-probdef}).

{\it (only if part)}:
Suppose system~(\ref{eqn:ineq-nz-system-probdef}) has a solution in $[0, 2 \SolBound]$.
This means that the original system~(\ref{eqn:ineq-system-probdef}) is feasible.
It follows that system~(\ref{eqn:ineq-nz-system-probdef}') has a solution in $[0, \SolBound]$.
\end{proof}
\end{enumerate}
In both solutions, we must search $2 \SolBound + 1$ values for 
each variable $x_j$, $1 \leq j \leq n$. However, the former avoids the need to
add $x_0$, and hence will have fewer input variables in the SAT-encoding.
Hence, the former solution is preferable.

The reader must note, though, that this optimization is only relevant
when the introduction of the zero variable (significantly) affects the
value of $\amax$. (The impact on $w$ is minor.) If the value of $\amax$ is
unaffected by the introduction of the zero variable $x_0$, using $x_0$
can result in a more compact SAT-encoding than using an enumeration
domain of $[-\SolBound, \SolBound]$ for each variable. If one uses the
$x_0$ variable, one introduces $\log \SolBound$ input Boolean variables for $x_0$ in 
the SAT-encoding. On the other hand, without the $x_0$ variable, one 
introduces $n$ additional Boolean variables to encode sign bits. The
relative size of the SAT-encoding, and hence the decision to introduce
$x_0$, would depend
on whether $n$ significantly exceeds $\log \SolBound$.

\subsubsection{Product of $k$ largest coefficients and widths}
\label{sec:large-coeff-simple}

There is a simpler optimization which we have found to be useful in practice.

In the proof of Theorem~\ref{thm:main-bound}, in deriving the $(\maxA \cdot w)^k$ term,
we have assumed the worst-case scenario of each term in the determinant expansion
equaling $\maxA^k$ and there being $w$ terms to choose from in each row.

In fact, we can replace $\maxA^k$ with
$\prod_{i=1}^{k} {\maxA}_i$, where ${\maxA}_i$ denotes the largest coefficient in row $i$, in absolute
value. Similarly, $w^k$ can be replaced with $\prod_i w_i$, where $w_i$ is the width of constraint 
$i$. 

\subsection{Dealing with Large Constant Terms}
\label{sec:large-constterm}

For some formulas, the value of $\maxB$ is very large due to the presence of a single large
constant (or very few of them). In such cases, a less conservative analysis or other
problem transformations are useful. We present two such techniques here.

\subsubsection{Product of $s$ largest constants}
\label{sec:large-constterm-simple}

It is easy to see that,
in the proof of Theorem~\ref{thm:main-bound}, the $s \, \maxB$ term can be replaced by 
$\sum_{j=1}^{s} |b_{i_j}|$, where $b_{i_1}, b_{i_2}, \ldots, b_{i_s}$ are the $s$ largest 
elements of $b$ in absolute value. 
Similarly, the expression for $\Delta$ derived in Theorem~\ref{thm:disj-bound} gets modified to
$$\Delta = \biggl( \sum_{j=1}^{s} (|b_{i_j}|+1) \biggr) \, \cdot \, (\maxA \, w)^k$$

Like the optimization of Section~\ref{sec:large-coeff-simple}, this 
has also proved fairly useful in practice.

\subsubsection{Shift of origin}
\label{sec:shift-of-origin}

Another transformation that can be useful for dealing with large constant terms
is to replace a variable $x_j$ by $x_j - \alpha_j$; this corresponds to shifting
the origin in $\mathbb{R}^n$ by $\alpha_j$ along the $x_j$-axis.

The $i^{\text{th}}$ 
constraint is then transformed into $\sum_j a_{i,j} (x_j - \alpha_j) \geq b_i$.
Rewriting this, we obtain the form
$\sum_j a_{i,j} x_j \geq b_i'$, where $b_i' = b_i + (\sum_j a_{i,j} \alpha_j)$.

The new value of $\maxB$, after the transformation, is $\max_i |b_i'|$.
Therefore, we wish to find values of $\alpha_j$s so as to minimize the value of $\max_i |b_i'|$.

This problem can be phrased as the following integer linear program:
\begin{eqnarray*}
& \min z  \\
{\text{subject to}} & \\
z & \geq & b_i + (\sum_j a_{i,j} \alpha_j) \hspace*{2cm} 1 \leq i \leq m \\
z & \geq & - b_i - (\sum_j a_{i,j} \alpha_j) \hspace*{1.75cm} 1 \leq i \leq m \\
z & \geq & 0 \\
& & z \in \mathbb{Z}, \; \; \; \alpha_j \in \mathbb{Z} \; {\text{for}} \; 1 \leq j \leq n
\end{eqnarray*}
This ILP has $n+1$ variables and $2 m + 1$ constraints (including the non-negativity
constraint on $z$). 

In fact, one can write one such ILP for each variable class, since they do not share
any variables or constraints. Then, the optimum value for each class will indicate
the new value of $\maxB$ to use for that class.

\section{Implementation and Experimental Results}
\label{sec:results}

\subsection{Implementation}
\label{sec:impl}

We used the bound derived in the previous section to implement a decision procedure
based on finite instantiation. 

The procedure starts by analyzing the formula to obtain parameters, and computes
the solution bound. We found that the optimizations of Section~\ref{sec:var-classes}, 
\ref{sec:spl-case-bounds}, and~\ref{sec:coeff-zero-var-opt} 
are always useful, especially since formulas tend to contain many
variables classes comprising of only difference constraints. Hence, our base-line
implementation always includes these optimizations. The impact of other optimizations
is reported in Section~\ref{sec:results-opts}.

Given the solution bound,
integer variables in the QFP formula are encoded as symbolic bit-vectors large enough to express
any integer value within the bound. Arithmetic operators are implemented as arbitrary-precision
bit-vector arithmetic operations. Equalities and inequalities over integer expressions 
are translated to corresponding relations over bit-vector expressions. 
The resulting Boolean formula is passed as input to a SAT solver.

We implemented our procedure as part of the UCLID verifier~\cite{uclid-www},
which is written in Moscow ML~\cite{moscow-ml-www}.
In our implementation we used the zChaff SAT solver~\cite{zchaff-www}
version 2004.5.13.
In the sequel, we will refer to our decision procedure as the ``UCLID'' procedure.

\subsection{Experimental Results}

We report here on a series of experiments we performed to evaluate our decision 
procedure against other theorem provers, as well as to assess the impact of
the various optimizations discussed in Section~\ref{sec:enhancements}.\footnote{Note: The results 
presented in this section are an updated version of those reported in the LICS'04 conference version.}

All experiments were performed on a Pentium-IV $2$ GHz machine with $1$ GB of RAM running Linux. A timeout
of $3600$ seconds ($1$ hour) was imposed on each run.

\subsubsection{Benchmarks}
\label{sec:benchmarks}

For benchmarks, we used $10$ formulas from
the Wisconsin Safety Analyzer (WiSA) project
on checking format string vulnerabilities, and $3$ generated by 
the Blast software model checker.
The benchmarks include both satisfiable and unsatisfiable
formulas in an extension of QFP with uninterpreted functions. Uninterpreted functions were first
eliminated using Ackermann's technique~\cite{ackermann-54},\footnote{Ackermann's function
elimination method replaces each function application by a fresh variable, and then instantiates
the congruence axiom for those applications. For instance, the formula $f(x) = f(y)$ is
translated to the function-free formula ${v_f}_1 = {v_f}_2 \land (x = y \implies {v_f}_1 = {v_f}_2)$.}  
and the decision procedures were run on the resulting QFP formula. 

Some characteristics of the formulas are displayed in 
Table~\ref{tbl:benchmarks}.
For each formula,
we indicate whether it is satisfiable or not. We give the values of parameters 
$n$, $m$, $k$, $w$, $\maxA$ and $\maxB$ corresponding to the variable class for which 
$S = \lceil \log [(n+2) \Delta] \rceil$
is largest, i.e, for which we need the largest number of bits per variable.
The values of the parameters for the overall formula are also given (although these
are not used in computing $S$ for any variable class); thus, the values of $m$ and $n$ in 
these columns are the 
{\em total} numbers of variables and constraints for the entire formula.

The top $10$ formulas listed in the table are from the WiSA project.
One key characteristic of these formulas is that they involve a significant number of 
Boolean operators ($\land$, $\lor$, $\neg$), and in particular there is a lot of
alternation of $\land$ and $\lor$. 
The other important characteristic of these benchmarks is that, although they
vary in $n$, $m$, and $\maxB$, the values of $k$, $w$, and $\maxA$ are fixed at a small value.

Three formulas from the Blast suite are listed at the bottom of Table~\ref{tbl:benchmarks}.
All these formulas are unsatisfiable.
Each formula is a conjunction of two sub-formulae: 
a large conjunction of linear constraints, and 
a conjunction of congruence constraints generated by Ackermann's function elimination
method. Thus, there is only one alternation of $\land$ and $\lor$ in these formulas.

\begin{table}[ht]
\begin{center}
\begin{tabular} {|c|c||r|r|r|r|r|r|r||r|r|r|r|r|r|r|}
\hline
Formula & Ans. &
\multicolumn{7}{|c||}{Parameters corr. to max. $S$} &
\multicolumn{6}{|c|}{Max. parameters overall} 
\\ \cline{3-9} \cline{10-15}
& & $n$ & $m$ & $k$ & $w$ & $\maxA$ & $\maxB$ & $S$ &
$n$ & $m$ & $k$ & $w$ & $\maxA$ & $\maxB$ 
\\ \hline

s-20-20 & SAT & 28 & 263 & 5 & 4 & 4 & 21 & 36 & 64 & 550 & 5 & 4 &4  & 255 \\ 

s-20-30 & SAT & 28 & 263 & 5 & 4 & 4 & 30 & 36 & 64 & 550 & 5 & 4 &4  & 255 \\ 

s-20-40 & UNS & 28 & 263 & 5 & 4 & 4 & 40 & 37 & 64 & 550 & 5 & 4 &4  & 255 \\ 

s-30-30 & SAT & 38 & 383 & 5 & 4 & 4 & 31 & 37 & 82 & 800 & 5 & 4 &4  & 255\\ 

s-30-40 & SAT & 38 & 383 & 5 & 4 & 4 & 40 & 37 & 82 & 800 & 5 & 4 &4  & 255\\ 

xs-20-20 & SAT & 49 & 323 & 5 & 4 & 4 & 21 & 37 & 84 & 632 & 5 & 4 & 4 & 255 \\ 

xs-20-30 & SAT & 49 & 323 & 5 & 4 & 4 & 30 & 38 & 84 & 632 & 5 & 4 & 4 & 255\\

xs-20-40 & UNS & 49 & 323 & 5 & 4 & 4 & 40 & 38 & 84 & 632 & 5 & 4 & 4 & 255\\ 

xs-30-30 & SAT & 69 & 473 & 5 & 4 & 4 & 31 & 39 & 114 & 922 & 5 & 4 & 4 & 255 \\ 

xs-30-40 & SAT & 69 & 473 & 5 & 4 & 4 & 40 & 39 & 114 & 922 & 5 & 4 & 4 & 255\\ 

\hline

blast-tl2 & UNS & 54 & 67 & 7 & 3 & 1 & 0 & 24 & 145 & 274 & 7 & 3 & 1 & 128\\ 

blast-tl3 & UNS & 201 & 2669 & 19 & 6 & 1 & 15 & 70 & 260 & 2986 & 19 & 6 & 1 & 128\\ 

blast-f8 & UNS & 255 & 6087 & 0 & 2 & 1 & 2560 & 20 & 321 & 7224 & 0 & 2 & 1 & 2560\\ 

\hline
\end{tabular}
\end{center}
\caption{{\bf{Benchmark characteristics.}} The top half of the table consists of the WiSA 
benchmarks and the bottom three are generated by the Blast software verifier.}
\label{tbl:benchmarks}
\end{table}

\subsubsection{Impact of optimizations}
\label{sec:results-opts}

\newcommand{\BaseEnc}{{\sf{Base}}\xspace}
\newcommand{\CoeffEnc}{{\sf{Coeff}}\xspace}
\newcommand{\ConstEnc}{{\sf{Const}}\xspace}
\newcommand{\AllEnc}{{\sf{All}}\xspace}

In this section, we discuss the impact of optimizations discussed in Sections~\ref{sec:large-coeff}
and~\ref{sec:large-constterm}.

Table~\ref{tbl:results-opts} compares the following
$4$ different encoding options based on different ways of computing the solution bound:
\begin{itemize}
\item[\BaseEnc:] The base-line method of computing the solution bound.
\item[\CoeffEnc:] Using the optimization of Section~\ref{sec:large-coeff-simple} alone.
\item[\ConstEnc:] Using the optimization of Section~\ref{sec:large-constterm-simple} alone.
\item[\AllEnc:] Using optimization methods of both Sections~\ref{sec:large-coeff-simple} 
and~\ref{sec:large-constterm-simple}.
\end{itemize}
The comparison is made with respect to the largest number of bits needed for any
variable class, and the run-times for both generating the SAT-encoding and for SAT solving.

\begin{table}[ht]
{\small{
\begin{center}
\begin{tabular} {|c|c||r|r|r|r||r|r|r|r||r|r|r|r|}
\hline
Formula & Ans. 
& \multicolumn{4}{|c||}{Max. \#bits/var.} 
& \multicolumn{4}{|c||}{Encoding Time (sec.)} 
& \multicolumn{4}{|c|}{SAT Time (sec.)} 
\\ \cline{3-6} \cline{7-10} \cline{11-14} 
& 
& \printSideWays{\BaseEnc} & \printSideWays{\CoeffEnc} & \printSideWays{\ConstEnc} & \printSideWays{\AllEnc} 
& \printSideWays{\BaseEnc} & \printSideWays{\CoeffEnc} & \printSideWays{\ConstEnc} & \printSideWays{\AllEnc} 
& \printSideWays{\BaseEnc} & \printSideWays{\CoeffEnc} & \printSideWays{\ConstEnc} & \printSideWays{\AllEnc} 
\\ \hline

s-20-20
 & SAT 
& 36
& 26
& 31
& 21
 & 1.26
 & 0.98
 & 1.12
 & 0.73
 & 0.27  
 & 0.27 
 & 0.20 
 & 0.20  
\\

s-20-30
 & SAT 
& 36
& 26
& 31
& 22
 & 1.29
 & 1.03
 & 1.05
 & 0.76
 & 0.38 
 & 0.57 
 & 0.41 
 & 0.36
\\

s-20-40
 & UNS 
& 37
& 27
& 32
& 22
 & 1.29
 & 0.99
 & 1.02 
 & 0.73 
 & 0.72
 & 0.61
 & 0.95
 & 0.39
\\

s-30-30
 & SAT 
& 37
& 27
& 32
& 22
 & 2.03
 & 1.41 
 & 1.48
 & 1.13
 & 1.55
 & 0.55
 & 0.26
 & 0.63 
\\

s-30-40
 & SAT 
& 37
& 28
& 32
& 23
 & 2.03
 & 1.48
 & 1.47
 & 1.13
 & 3.03
 & 2.10 
 & 0.41
 & 1.08
\\

xs-20-20
 & SAT 
& 37
& 28
& 32
& 22
 & 1.89
 & 1.36
 & 1.40
 & 1.04
 & 0.51
 & 0.55
 & 0.97
 & 0.31
\\

xs-20-30
 & SAT 
& 38
& 28
& 32
& 23
 & 1.94
 & 1.31 
 & 1.68
 & 1.08  
 & 1.09
 & 1.85 
 & 1.00
 & 0.69 
\\

xs-20-40
 & UNS 
& 38
& 29
& 33
& 23
 & 1.91
 & 1.42
 & 1.55
 & 1.09
 & 4.45
 & 4.41
 & 3.90
 & 2.80
\\

xs-30-30
 & SAT 
& 39
& 29
& 33
& 23
 & 2.89
 & 2.32 
 & 2.48
 & 1.57 
 & 2.91
 & 4.29
 & 0.78
 & 0.88 
\\

xs-30-40
 & SAT 
& 39
& 30
& 33
& 24
 & 2.86
 & 2.36 
 & 2.67
 & 1.61 
 & 1.61
 & 2.88 
 & 0.92
 & 1.55
\\

\hline

blast-tl2
 & UNS
& 24
& 24
& 19
& 19
 & 0.65
 & 0.65
 & 0.50
 & 0.50
 & 0.02 
 & 0.02 
 & 0.02
 & 0.01 
\\

blast-tl3
 & UNS
& 70
& 53
& 62
& 46
 & 29.20
 & 19.12
 & 22.29
 & 16.94
 & 0.82
 & 0.62 
 & 0.66
 & 0.49
\\

blast-f8
 & UNS
& 20
& 20
& 12
& 12
 & 17.54
 & 17.56
 & 10.37
 & 10.36
 & 2.02 
 & 2.02 
 & 0.96
 & 0.96
\\

\hline
\end{tabular}
\end{center}
\caption{{\bf{An experimental evaluation of encoding optimizations.}} We compare the $4$ different
UCLID encoding options with respect to the maximum number of bits needed for any integer 
variable (``Max. \#bits/var.''), 
the time taken to generate the Boolean encoding, and the time
taken by the SAT solver. 
}
\label{tbl:results-opts}
}} 
\end{table}

First, we note that \CoeffEnc and \ConstEnc both generate more compact encodings than \BaseEnc;
on the WiSA benchmarks, they use about $5$-$10$ fewer bits per variable in the largest
variable class. The reduction in the total number of bits, summed over all variables in 
all variable classes, is similar, since most variables fall into a single class. 

The encoding times decrease with reduction in number of bits; this is just as one would predict.

However, the comparison of SAT solving times is more mixed; on a few benchmarks
\CoeffEnc and \ConstEnc outperform \BaseEnc, and on others, they do worse. The latter behavior is
observed especially on satisfiable formulas. The reason for this appears to be a relative ease
in finding larger solutions for those formulas than finding smaller solutions.

When \CoeffEnc and \ConstEnc are both used (indicated as ``\AllEnc''), we find that not only
are encoding times smaller than the \BaseEnc technique, but SAT solving times are also
smaller in all cases. 
This seems to indicate that a reduction in SAT-encoding size beyond a certain limit overcomes
any negative effects of restricting the search to smaller solutions. 

We also performed an experiment to explore the use of the {\em shift-of-origin} optimization
described in Section~\ref{sec:shift-of-origin}. UCLID automatically formulated
the ILP and solved it using the CPLEX optimization tool~\cite{cplex-www} (version 8.1).
Since none of the benchmarks listed
in Table~\ref{tbl:benchmarks} have especially large constants, we used a different,
unsatisfiable formula from the Blast suite which has only difference constraints, but
with large constants.

Table~\ref{tbl:shift-of-origin} summarizes the key characteristics of this formula
as well as the results obtained by comparing versions of the base-line (\BaseEnc) implementation
with and without the optimization enabled. We list the values of parameters, with and without
the shift-of-origin optimization enabled, for the variable
classes that yield the two largest values of $S$ when the optimization is disabled.
\begin{table}[ht]
\begin{center}
\begin{tabular} {|c||r|r|r|r||r|r|r|r||r||r|r|r|}
\hline
Shift-of-origin &
\multicolumn{4}{|c||}{Param. for largest $S$} &
\multicolumn{4}{|c||}{Param. for $2^{\text{nd}}$ largest $S$} &
Total &
\multicolumn{2}{|c|}{Time (sec.)} 
\\ \cline{2-5} \cline{6-9} \cline{11-12}
enabled? &
$n$ & $m$ & $\maxB$ & $S$ & 
$n$ & $m$ & $\maxB$ & $S$ & 
\#bits &
Enc. & SAT 
\\ \hline

No & 230 & 6417 & 2162688 & 29 & 2 & 2 & 261133242 & 28 & 7510 
& 24.68 & 0.70 \\ \hline
Yes & 230 & 6417 & 432539 & 27 & 2 & 2 & 0 & 1 & 6833
& 25.78 & 0.71 \\
\hline
\end{tabular}
\end{center}
\caption{{\bf{Evaluating the shift-of-origin optimization.}} 
We list the values of parameters corresponding to variable classes with
the two largest values of $S$, as computed {\em without} the shift-of-origin optimization.
``Total \#bits'' indicates the number of bits needed to encode all integer variables.
Encoding time is indicated as ``Enc.''
and SAT solving time as ``SAT''.
}
\label{tbl:shift-of-origin}
\end{table}

With the optimization turned on, the largest constant in the {\em entire} formula falls from
$261133242$ to $432539$, a $600$-fold reduction. However, if we restrict our attention 
to the largest variable class, comprising $230$ variables, the reduction in $\maxB$ is more modest,
about a factor of $4$.
This yields a saving of $2$ bits per variable for that variable class. The saving in the total
number of bits, summed over all variable classes, is $677$. This is, however, not large
enough to reduce either the encoding time or the SAT time. In fact, the encoding
time increases by about a second; this is the time required to run CPLEX and for the processing overhead
of creating the ILP.

Even though the shift-of-origin optimization has not resulted in faster run-times in our
experiments, it clearly has the potential to greatly reduce the number of bits,
and might prove useful on other benchmarks.

\subsubsection{Comparison with other theorem provers}

We compared UCLID's performance with that of the SAT-based provers 
ICS~\cite{ics-www} (version 2.0) and CVC-Lite~\cite{cvcl-www} (the new implementation
of CVC, version 2.0.0),\footnote{Note that the results for CVC-Lite 2.0.0 are a significant
improvement over those we previously obtained~\cite{seshia-thesis05} using an older version.}
as well as the automata-based procedure LASH~\cite{lash-www} (version 0.9).
While CVC-Lite and LASH are sound and complete for QFP, ICS 2.0 is incomplete;
i.e., it can report a formula to be satisfiable when it is not. The 
ground decision procedure ICS uses is the Simplex linear 
programming algorithm with some additional heuristics to deal with 
integer variables.
However, in our experiments, both UCLID and ICS returned the same answer
whenever ICS terminated within the timeout.
The ground decision procedure for CVC-Lite is a proof-producing 
variant of the Omega test~\cite{ganesh-tacas03}.

LASH was unable to complete on any benchmark within the timeout since it was
unable to construct the corresponding automaton; we attribute
this to the relatively large number of variables and constraints in our formulas, and
note that Ganesh et al. obtained similar results in their study~\cite{ganesh-fmcad02}.

\begin{table}[ht]
\begin{center}
\begin{tabular} {|c|c||r|r|r||r|r|r||r|}
\hline
Formula & Ans. &
\multicolumn{3}{|c||}{UCLID Time} &
\multicolumn{3}{c||}{ICS} 
& \multicolumn{1}{c|}{CVC-Lite}
\\ \cline{6-9}
& 
& \multicolumn{3}{|c||}{(sec.)} 
& \#(Inc. & \multicolumn{2}{|c||}{Time (sec.)} 
& Total Time 
\\ \cline{3-5} \cline{7-8}
& 
& Enc. & SAT & Total
& assn.) & Ground & Total & (sec.) 
\\ \hline
s-20-20 & SAT &  0.73 & 0.20 & 0.93 & 904 & 23.32 & 23.76 & 1.45  \\ 

s-20-30 & SAT &  0.76 & 0.36 & 1.12 & 1887 & 51.68 & 52.29  & 1.73 \\ 

s-20-40 & UNS &  0.73 & 0.39 & 1.12 & 25776 & 658.01 & 669.99  & *  \\ 

s-30-30 & SAT &  1.13 & 0.63 & 1.76 & 2286 & 268.21 & 269.42  & 3.83 \\ 

s-30-40 & SAT &  1.13 & 1.08 & 2.21 & 14604 & 1621.27 & 1625.15  & 4.28\\

xs-20-20 & SAT & 1.04 & 0.31 & 1.35 & 2307 & 97.21 & 98.32  & 1.78  \\ 

xs-20-30 & SAT & 1.08 & 0.69 & 1.77 & 33103 & 1519.77 & 1540.27  & 2.04 \\ 

xs-20-40 & UNS & 1.09 & 2.80 & 3.89 & 97427 & 3468.91 & *  & *\\ 

xs-30-30 & SAT & 1.57 & 0.88 & 2.45 & 72585 & 3287.47 & *  & 4.90 \\ 

xs-30-40 & SAT & 1.61 & 1.55 & 3.16 & 33754 & 3082.34 & *  & 4.36 \\ 

\hline

blast-tl2 & UNS & 0.50 & 0.01 & 0.51 & 1 & 0.01 & 0.01  & 0.15 \\ 

blast-tl3 & UNS & 16.94 & 0.49 & 17.43 & 0 & 0.00 & 0.01 & 2.66 \\ 

blast-f8 & UNS & 10.36 & 0.96 & 11.32 & 1 & 0.01 & 0.05 & 14.55 \\

\hline
\end{tabular}
\end{center}
\caption{{\bf{Experimental comparison with other theorem provers.}}
The UCLID version is the one with all optimizations turned on (``\AllEnc'').
For ICS, we give the total time, the number of inconsistent Boolean assignments analyzed by the ground decision
procedure (``\#(Inc. assn.)''), 
as well as the overall time taken by the ground decision procedure (``Ground'').
For CVC-Lite, we indicate the total run-time.
A ``$*$'' indicates that the decision procedure timed out after $3600$ sec.
LASH did not complete within the timeout on any formula.
}
\label{tbl:results}
\end{table}

A comparison of UCLID versus ICS and CVC-Lite is displayed in Table~\ref{tbl:results}. 
From Table~\ref{tbl:results}, we observe that
UCLID outperforms ICS on all the WiSA benchmarks, terminating within a few seconds on each one.
However, ICS performs best on the Blast formulas, finishing within a fraction of a second
on all. CVC-Lite runs much faster than ICS on the satisfiable WiSA formulas, but does not
finish on either of the unsatisfiable WiSA formulas, and does not outperform UCLID on any of the
WiSA benchmarks.
However, it outperforms UCLID on one of the Blast formulas. 
Due to the unavailability of statistics on where CVC-Lite spends its time,
we can only present a detailed comparison between UCLID and ICS here. 
We believe that CVC-Lite's superior performance to ICS on satisfiable formulas
is mainly due to improved Boolean simplification heuristics and, to a lesser extent,
due to a faster ground decision procedure.\footnote{Based on personal communication with
S. Berezin.} 
The better performance
compared to UCLID on one of the Blast formulas is because that formula is propositionally
unsatisfiable, as we will discuss in more detail below.

Let us consider the WiSA benchmarks first.
These formulas have a non-trivial Boolean structure
that requires ICS to enumerate many inconsistent Boolean assignments before
being able to decide the formula.
The ICS run-time is dominated by the time taken by the ground decision procedure.
We observe that the number of inconsistent Boolean assignments alone is not a precise indicator of
total run-time, which also depends on the time taken by the ground decision procedure in 
ruling out a single Boolean assignment. Further optimization of ICS's ground decision procedure
might improve its overall run-time, at least on the satisfiable formulas.\\
$\indent$ 
The reason for UCLID's superior performance 
is the formula structure, where $k$, $w$, and $\maxA$ remain
fixed at a low value while $m$, $n$, and $\maxB$ increase. Thus, the maximum number of bits 
per variable stays about the same even as $m$ increases substantially, 
and the resulting SAT problem is within the capacity of zChaff.
The times for both encoding and SAT solving phases are small.
In particular, the small SAT solving time on the unsatisfiable instances 
indicates that the proof of unsatisfiability is also small.

Next, consider the results on the Blast formulas.
The reason for ICS's superior performance on these can be gauged
by the number of inconsistent Boolean assignments it has to enumerate. On
the formula named ``blast-tl3'', purely Boolean reasoning suffices to decide
unsatisfiability. For the other two formulas, the reason for unsatisfiability
is a mutually-inconsistent subset amongst all the
linear constraints that are conjoined together, and a single
call to ICS's ground decision procedure suffices to infer the inconsistency. 
In all three cases, the ``proof of unsatisfiability'' that ICS must find
is small.
\\
$\indent$ 
On the other hand, UCLID's run-time is dominated by the encoding time. Once the
encoding is generated, the SAT solver decides unsatisfiability easily.

To summarize, it appears that decision procedures like ICS and CVC-Lite, which are based 
on a lazy translation to SAT, are effective when the formula structure is such that only 
a few calls to the ground decision procedure are required (i.e., satisfiable solutions are
easy to find, or the proof of unsatisfiability is shallow), and the ground decision procedure 
is itself efficient. UCLID performs better 
on formulas with complicated Boolean structure and comprising linear constraints
with the sparse structure formalized in this paper.

\section{Conclusions and Future Work}
\label{sec:concl}
In this paper, we have presented a formal approach to exploiting 
the ``sparse, mainly difference constraint'' nature of quantifier-free
Presburger formulas 
encountered in software verification. Our approach is based on formalizing
this sparse structure using new parameters, and deriving a new
parameterized bound on satisfying solutions to QFP formulas. We have also proposed
several ways in which the bound can be reduced in practice. Experimental
results show the benefits of using the derived bound in a
SAT-based decision procedure based on finite instantiation.

The work described in this paper can be extended in a few
new directions.
Some of these are discussed below.

\subsection{Computing the Solution Bound Lazily}
\label{sec:lazy}
In our implementation, we compute a conservative bound and translate a QFP formula
to a Boolean formula in a single step.
An alternative approach is to perform this transformation {\it lazily}, increasing
the solution bound ``on demand''.

One such lazy encoding approach works, in brief, as follows. (Details can be found
in the paper by Kroening et al.~\cite{kroening-cav04}.) 

We start with an encoding size for each integer variable
that is smaller than that prescribed by the conservative bound (say, $1$ bit per variable).

If the resulting Boolean formula is satisfiable,
so is the original QFP formula. 
If not, the proof of unsatisfiability generated by the 
SAT solver is used to generate a sound abstraction of the original formula, which can be checked with
a sound and complete decision procedure for QFP (such as the one proposed
in this paper). If this decision procedure concludes that the abstraction is unsatisfiable,
so is the original formula, but if not, it provides a counterexample which indicates the necessary
increase in the encoding size. A new SAT-encoding is generated, 
and the procedure repeats. 

The bound $S$ on solution size that we derive in this paper implies an
upper bound $n S$ on the number of iterations of this lazy encoding procedure; thus
the lazy encoding procedure needs only polynomially many iterations before 
it terminates with the correct answer.

The potential advantage of this lazy approach is two-fold: (1) It avoids using the 
conservative bounds we have derived in this paper, and (2) if the generated 
abstractions are small, the sound and complete decision procedure 
used by this approach will run much faster than if it were fed the original
formula. 

For the WiSA benchmarks discussed in Section~\ref{sec:results}, we found that
a solution bound of $2^8 - 1$, i.e., $8$ bits per variable, is sufficient to 
decide satisfiability. However, the time required to derive this bound using
the method of~\cite{kroening-cav04} is much greater than the run-times we report in
Section~\ref{sec:results}. 
Still, the lazy approach might prove especially useful 
in cases in which $S$ is so large that the SAT problem is outside the
reach of current SAT solvers.

\subsection{Special Classes of Constraints}

In Section~\ref{sec:spl-case-bounds}, we saw that if all linear constraints are 
difference constraints, a tighter solution bound can be used. Recently, we
have derived a tighter bound for a special class of constraints that is a 
superset of difference constraints.
Constraints in this class refer to at most two variables ($w = 2$), and all
variable coefficients are in $\{0, -1, +1\}$ (i.e., $\maxA \leq 1$).
These constraints are referred to in literature as either 
{\em generalized 2SAT constraints} or {\em unit two-variable per inequality} constraints.
For this special case, we have derived a solution bound of 
$2 \cdot \min(n,m) \cdot (\maxB + 1)$~\cite{seshia-tr04},
exactly twice the bound for difference logic. The proof techniques for
deriving this bound are quite different from those used in this paper.

It would be interesting to find other special constraint classes for which
the bounds presented in this paper can be further tightened.

\subsection{Other Directions and Open Problems}

As we have observed in Section~\ref{sec:results}, the impact of reduction
of number of bits on the SAT solving time is not always predictable. We
are currently trying to better understand the reasons for this.

Encoding to SAT is not the only way in which the bounds presented in 
this paper can be used. It would be interesting to explore non-SAT-based
decision procedures based on the bounds we derive.

The theoretical results of this paper rest heavily on the 
bound $(n+2) \cdot \Delta$ given by Borosh, Treybig, and Flahive, stated
in Theorem~\ref{thm:borosh-treybig}. 
In their 1992 paper~\cite{borosh-siam-mtx92}, Borosh and Treybig conjectured
that this bound can be improved to just $\Delta$. To our knowledge,
this conjecture is still open.

Finally, it would also be interesting to apply our work to
areas outside of software verification that share the special
structure of linear constraints exploited in this paper.


\section*{Acknowledgments}
We are grateful to Jo{\"e}l Ouaknine for his inputs and a careful
reading of the proofs, and to Ofer Strichman and K. Subramani for
valuable discussions.
We thank Sagar Chaki, Michael Ernst, Vinod Ganapathy, 
Somesh Jha, Ranjit Jhala, and Stephen McCamant 
for providing us with benchmark formulas. 
We also thank Sergey Berezin, Leonardo de Moura, and Louis Latour for help with 
CVC-Lite, ICS, and LASH respectively. This research was supported in part
by ARO grant DAAD19-01-1-0485.


\begin{thebibliography}{MMZ{\etalchar{+}}01}

\bibitem[Ack54]{ackermann-54}
W.~Ackermann.
\newblock {\em Solvable Cases of the Decision Problem}.
\newblock North-Holland, Amsterdam, 1954.

\bibitem[BC72]{brenner-amm72}
Joel Brenner and Larry Cummings.
\newblock The {Hadamard} maximum determinant problem.
\newblock {\em American Mathematical Monthly}, 79:626--630, June-July 1972.

\bibitem[BD02]{brinkmann-vlsi02}
R.~Brinkmann and R.~Drechsler.
\newblock {RTL}-datapath verification using integer linear programming.
\newblock In {\em Proceedings of the IEEE {VLSI} Design Conference}, pages
  741--746, 2002.

\bibitem[BDS02]{barrett-cav02}
C.~Barrett, D.~L. Dill, and A.~Stump.
\newblock Checking satisfiability of first-order formulas by incremental
  translation to {SAT}.
\newblock In E.~Brinksma and K.~G. Larsen, editors, {\em Proc. 14th Intl.
  Conference on Computer-Aided Verification ({CAV}'02)}, LNCS 2404, pages
  236--249. Springer-Verlag, July 2002.

\bibitem[BFRT89]{borosh-pams89}
I.~Borosh, M.~Flahive, D.~Rubin, and L.~B. Treybig.
\newblock A sharp bound for solutions of linear {Diophantine} equations.
\newblock {\em Proceedings of the American Mathematical Society},
  105(4):844--846, April 1989.

\bibitem[BFT86]{borosh-dismath86}
I.~Borosh, M.~Flahive, and L.~B. Treybig.
\newblock Small solutions of linear {Diophantine} equations.
\newblock {\em Discrete Mathematics}, 58:215--220, 1986.

\bibitem[BGD03]{ganesh-tacas03}
S.~Berezin, V.~Ganesh, and D.~L. Dill.
\newblock An online proof-producing decision procedure for mixed-integer linear
  arithmetic.
\newblock In {\em Proc. Tools and Algorithms for the Construction and Analysis
  of Systems (TACAS'03)}, LNCS 2619, pages 521--536, April 2003.

\bibitem[BLS02]{bryant-cav02}
R.~E. Bryant, S.~K. Lahiri, and S.~A. Seshia.
\newblock Modeling and verifying systems using a logic of counter arithmetic
  with lambda expressions and uninterpreted functions.
\newblock In E.~Brinksma and K.~G. Larsen, editors, {\em Proc. Computer-Aided
  Verification ({CAV}'02)}, LNCS 2404, pages 78--92, July 2002.

\bibitem[BT76]{borosh-pams76}
I.~Borosh and L.~B. Treybig.
\newblock Bounds on positive integral solutions of linear {Diophantine}
  equations.
\newblock {\em Proceedings of the American Mathematical Society},
  55(2):299--304, March 1976.

\bibitem[BT92]{borosh-siam-mtx92}
I.~Borosh and L.~B. Treybig.
\newblock A sharp bound on positive solutions of linear {Diophantine}
  equations.
\newblock {\em SIAM Journal on Matrix Analysis and Applications},
  13(2):454--458, April 1992.

\bibitem[CCG{\etalchar{+}}03]{chaki-icse03}
Sagar Chaki, Edmund~M. Clarke, Alex Groce, Somesh Jha, and Helmut Veith.
\newblock Modular verification of software components in {C}.
\newblock In {\em Proc. 25th International Conference on Software Engineering
  (ICSE)}, pages 385--395, 2003.

\bibitem[Cha93]{chandru-compjnl93}
Vijay Chandru.
\newblock Variable elimination in linear constraints.
\newblock {\em The Computer Journal}, 36(5):463--472, August 1993.

\bibitem[{CPL}]{cplex-www}
{CPLEX Optimization Tool.}
\newblock {Available from ILOG. {\tt http://www.ilog.com/products/cplex/}}.

\bibitem[{{CVC}}]{cvcl-www}
{{CVC-Lite}: Cooperating Validity Checker}.
\newblock {Available at {\tt http://verify.stanford.edu/CVCL/}}.

\bibitem[dMRS02]{demoura-cade02}
Leonardo de~Moura, Harald Rue{\ss}, and Maria Sorea.
\newblock Lazy theorem proving for bounded model checking over infinite
  domains.
\newblock In {\em Proc. 18th International Conference on Automated Deduction
  (CADE)}, pages 438--455, 2002.

\bibitem[DNS03]{detlefs-tr03}
David Detlefs, Greg Nelson, and James~B. Saxe.
\newblock Simplify: {A} theorem prover for program checking.
\newblock Technical Report HPL-2003-148, HP Laboratories Palo Alto, 2003.

\bibitem[FR74]{fischer-siam74}
M.~J. Fischer and M.~O. Rabin.
\newblock Super-exponential complexity of {P}resburger arithmetic.
\newblock {\em {Proceedings of SIAM-AMS}}, 7:27--41, 1974.

\bibitem[GBD02]{ganesh-fmcad02}
V.~Ganesh, S.~Berezin, and D.~L. Dill.
\newblock Deciding {P}resburger arithmetic by model checking and comparisons
  with other methods.
\newblock In {\em {Formal Methods in Computer-Aided Design (FMCAD '02)}}, LNCS
  2517, pages 171--186. Springer-Verlag, November 2002.

\bibitem[GN02]{goldberg-date02}
E.~Goldberg and Y.~Novikov.
\newblock {BerkMin}: {A} fast and robust {SAT} solver.
\newblock In {\em Design Automation and Test in Europe (DATE) 2002}, pages
  142--149, 2002.

\bibitem[HJMS02]{henzinger-popl02}
Thomas~A. Henzinger, Ranjit Jhala, Rupak Majumdar, and Gregoire Sutre.
\newblock Lazy abstraction.
\newblock In {\em Proc. 29th ACM Symposium on Principles of Programming
  Languages}, pages 58--70, 2002.

\bibitem[{{ICS}}]{ics-www}
{{ICS}: Integrated Canonizer and Solver}.
\newblock {Available at {\tt http://www.icansolve.com}}.

\bibitem[KM78]{kannan-oor78}
R.~Kannan and C.~L. Monma.
\newblock On the computational complexity of integer programming problems.
\newblock In {\em Optimisation and Operations Research}, volume 157 of {\em
  Lecture Notes in Economics and Mathematical Systems}, pages 161--172.
  Springer-Verlag, 1978.

\bibitem[KOSS04]{kroening-cav04}
Daniel Kroening, Jo{\"e}l Ouaknine, Sanjit~A. Seshia, and Ofer Strichman.
\newblock Abstraction-based satisfiability solving of {P}resburger arithmetic.
\newblock In {\em Proc. 16th International Conference on Computer-Aided
  Verification ({CAV})}, pages 308--320, July 2004.

\bibitem[{{LAS}}]{lash-www}
{{LASH} Toolset}.
\newblock {Available at {\tt http://www.
  montefiore.ulg.ac.be/\~{}boigelot/research/lash}}.

\bibitem[ME03]{mccamant-fse03}
Stephen McCamant and Michael~D. Ernst.
\newblock Predicting problems caused by component upgrades.
\newblock In {\em Proceedings of the 11th ACM SIGSOFT Symposium on Foundations
  of Software Engineering (FSE)}, pages 287--296, 2003.

\bibitem[MMZ{\etalchar{+}}01]{moskewicz-dac01}
M.~Moskewicz, C.~Madigan, Y.~Zhao, L.~Zhang, and S.~Malik.
\newblock Chaff: Engineering an efficient {SAT} solver.
\newblock In {\em {38th Design Automation Conference (DAC '01)}}, pages
  530--535, June 2001.

\bibitem[{Mos}]{moscow-ml-www}
{Moscow ML}.
\newblock {Available at {\tt http://www.dina.dk/ \verb|~|sestoft/mosml.html}}.

\bibitem[NO79]{nelson-toplas79}
G.~Nelson and D.~C. Oppen.
\newblock Simplification by cooperating decision procedures.
\newblock {\em {ACM Transactions on Programming Languages and Systems}},
  1(2):245--257, 1979.

\bibitem[Pap81]{papadim-jacm81}
Christos~H. Papadimitriou.
\newblock On the complexity of integer programming.
\newblock {\em Journal of the {ACM}}, 28(4):765--768, 1981.

\bibitem[Pra77]{pratt-77}
Vaughan Pratt.
\newblock Two easy theories whose combination is hard.
\newblock Technical report, Massachusetts Institute of Technology, 1977.
\newblock Cambridge, MA.

\bibitem[Pre29]{presburger-27}
M.~Pre{\ss}burger.
\newblock {\"{U}}ber die {V}ollst\"andigkeit eines gewissen {S}ystems der
  {A}rithmetik ganzer {Z}ahlen, in welchem die {A}ddition als einzige
  {O}peration hervortritt.
\newblock {\em Comptes-rendus du Premier Congr\`es des Math\'ematiciens des
  Pays Slaves}, 395:92--101, 1929.

\bibitem[PRSS99]{pnueli-cav99}
A.~Pnueli, Y.~Rodeh, O.~Shtrichman, and M.~Siegel.
\newblock Deciding equality formulas by small-domain instantiations.
\newblock In N.~Halbwachs and D.~Peled, editors, {\em Computer-Aided
  Verification}, volume 1633 of {\em Lecture Notes in Computer Science}, pages
  455--469. Springer-Verlag, July 1999.

\bibitem[PS82]{papadim-steiglitz-82ch13}
Christos~H. Papadimitriou and Kenneth Steiglitz.
\newblock {\em Combinatorial Optimization: Algorithms and Complexity},
  chapter~13.
\newblock Prentice-Hall, 1982.

\bibitem[Pug91]{pugh-sc91}
William Pugh.
\newblock The omega test: {A} fast and practical integer programming algorithm
  for dependence analysis.
\newblock In {\em Supercomputing}, pages 4--13, 1991.

\bibitem[Sch86]{schrijver-86}
Alexander Schrijver.
\newblock {\em Theory of Linear and Integer Programming}.
\newblock John Wiley and Sons, 1986.

\bibitem[Ses05]{seshia-thesis05}
Sanjit~A. Seshia.
\newblock {\em Adaptive Eager Boolean Encoding for Arithmetic Reasoning in
  Verification}.
\newblock PhD thesis, Carnegie Mellon University, 2005.

\bibitem[Sho84]{shostak-jacm84}
R.~E. Shostak.
\newblock Deciding combinations of theories.
\newblock {\em {Journal of the ACM}}, 31(1):1--12, 1984.

\bibitem[SR02]{shankar-rta02}
Natarajan Shankar and Harald Rue{\ss}.
\newblock Combining {Shostak} theories.
\newblock In Sophie Tison, editor, {\em Proc. Rewriting Techniques and
  Applications}, LNCS 2378, pages 1--18. Springer-Verlag, July 2002.

\bibitem[SSB02]{strichman-cav02}
O.~Strichman, S.~A. Seshia, and R.~E. Bryant.
\newblock Deciding separation formulas with {SAT}.
\newblock In E.~Brinksma and K.~G. Larsen, editors, {\em Proc. 14th Intl.
  Conference on Computer-Aided Verification ({CAV}'02)}, LNCS 2404, pages
  209--222. Springer-Verlag, July 2002.

\bibitem[SSB04]{seshia-tr04}
Sanjit~A. Seshia, K.~Subramani, and Randal~E. Bryant.
\newblock On solving {Boolean} combinations of generalized {2SAT} constraints.
\newblock Technical Report {CMU-CS-04-179}, Carnegie Mellon University, 2004.

\bibitem[Str02]{strichman-fmcad02}
O.~Strichman.
\newblock On solving {P}resburger and linear arithmetic with {SAT}.
\newblock In {\em {Formal Methods in Computer-Aided Design (FMCAD '02)}}, LNCS
  2517, pages 160--170. Springer-Verlag, November 2002.

\bibitem[{{UCL}}]{uclid-www}
{{UCLID} Verification System.}
\newblock {Available at {\tt http://www.cs.cmu.edu/\verb|~|uclid}}.

\bibitem[vzGS78]{gathen-pams78}
J.~von~zur Gathen and M.~Sieveking.
\newblock A bound on solutions of linear integer equalities and inequalities.
\newblock {\em Proceedings of the American Mathematical Society},
  72(1):155--158, October 1978.

\bibitem[WB95]{wolper-sas95}
Pierre Wolper and Bernard Boigelot.
\newblock An automata-theoretic approach to {Presburger} arithmetic
  constraints.
\newblock In {\em Proc. Static Analysis Symposium}, LNCS 983, pages 21--32,
  September 1995.

\bibitem[{{W}is}]{wisa-www}
{{W}isconsin Safety Analyzer Project}.
\newblock {\tt {http://www.cs.wisc.edu/wisa}}.

\bibitem[WW99]{wolfman-ijcai99}
Steven~A. Wolfman and Daniel~S. Weld.
\newblock The {LPSAT} engine and its application to resource planning.
\newblock In {\em Proceedings of the International Joint Conference in
  Artificial Intelligence (IJCAI)}, pages 310--317, 1999.

\bibitem[{zCh}]{zchaff-www}
{zChaff {B}oolean Satisfiability Solver.}
\newblock {Available at {\tt http:/\!/\!ee.princeton.edu/\~{}\!chaff/\!zchaff.php}}.

\end{thebibliography}
\newcommand{\etalchar}[1]{$^{#1}$}


\end{document}